\documentclass[10pt, conference, compsocconf]{IEEEtran}

\ifCLASSINFOpdf
  \usepackage[pdftex]{graphicx}
  \graphicspath{{../pdf/}{../jpeg/}}
  \DeclareGraphicsExtensions{.pdf,.jpeg,.png}
\else
  \usepackage[dvips]{graphicx}
  \graphicspath{{../eps/}}
  \DeclareGraphicsExtensions{.eps}
\fi
\usepackage{algorithm}
\usepackage{algpseudocode}
\usepackage[tight,footnotesize]{subfigure}
\usepackage{listings}
\usepackage[dvipsnames]{xcolor}
\usepackage{float}
\usepackage{pifont}
\usepackage{cite}
\usepackage{url}
\usepackage{color, colortbl}
\usepackage{xspace}
\usepackage{booktabs}
\usepackage[american]{babel}
\usepackage{listings}
\usepackage{flushend}
\usepackage{multirow}
\usepackage{amsmath}

\makeatletter
\def\therule{\makebox[\algorithmicindent][l]{\hspace*{.5em}\vrule height .75\baselineskip depth .25\baselineskip}}%

\newtoks\therules
\therules={}
\def\appendto#1#2{\expandafter#1\expandafter{\the#1#2}}
\def\gobblefirst#1{
  #1\expandafter\expandafter\expandafter{\expandafter\@gobble\the#1}}%
\def\LState{\State\unskip\the\therules}
\def\pushindent{\appendto\therules\therule}%
\def\popindent{\gobblefirst\therules}%
\def\printindent{\unskip\the\therules}%
\def\printandpush{\printindent\pushindent}%
\def\popandprint{\popindent\printindent}%

\algdef{SE}[WHILE]{While}{EndWhile}[1]
  {\printandpush\algorithmicwhile\ #1\ \algorithmicdo}
  {\popandprint\algorithmicend\ \algorithmicwhile}%
\algdef{SE}[FOR]{For}{EndFor}[1]
  {\printandpush\algorithmicfor\ #1\ \algorithmicdo}
  {\popandprint\algorithmicend\ \algorithmicfor}%
\algdef{S}[FOR]{ForAll}[1]
  {\printindent\algorithmicforall\ #1\ \algorithmicdo}%
\algdef{SE}[LOOP]{Loop}{EndLoop}
  {\printandpush\algorithmicloop}
  {\popandprint\algorithmicend\ \algorithmicloop}%
\algdef{SE}[REPEAT]{Repeat}{Until}
  {\printandpush\algorithmicrepeat}[1]
  {\popandprint\algorithmicuntil\ #1}%
\algdef{SE}[IF]{If}{EndIf}[1]
  {\printandpush\algorithmicif\ #1\ \algorithmicthen}
  {\popandprint\algorithmicend\ \algorithmicif}%
\algdef{C}[IF]{IF}{ElsIf}[1]
  {\popandprint\pushindent\algorithmicelse\ \algorithmicif\ #1\ \algorithmicthen}%
\algdef{Ce}[ELSE]{IF}{Else}{EndIf}
  {\popandprint\pushindent\algorithmicelse}%
\algdef{SE}[PROCEDURE]{Procedure}{EndProcedure}[2]
   {\printandpush\algorithmicprocedure\ \textproc{#1}\ifthenelse{\equal{#2}{}}{}{(#2)}}%
   {\popandprint\algorithmicend\ \algorithmicprocedure}%
\algdef{SE}[FUNCTION]{Function}{EndFunction}[2]
   {\printandpush\algorithmicfunction\ \textproc{#1}\ifthenelse{\equal{#2}{}}{}{(#2)}}%
   {\popandprint\algorithmicend\ \algorithmicfunction}%

\newcommand\etal{\emph{et al.}\xspace}
\newcommand\Modela{\emph{Liu et al.}\xspace}
\newcommand\Modelb{\emph{Werkhoven et al.}\xspace}

\newcommand\cparagraph[1]{\vspace{1.5mm}\noindent \textbf{#1}}

\definecolor{mygreen}{rgb}{0,0.6,0}
\newcommand{\SVMs} {\texttt{SVMs}\xspace}
\newcommand{\SVM} {\texttt{SVM}\xspace}
\newcommand{\PCA} {\texttt{PCA}\xspace}
\newcommand{\KNN} {\texttt{KNN}\xspace}
\newcommand{\ANN} {\texttt{ANN}\xspace}
\definecolor{Gray}{gray}{0.9}

\makeatother

\begin{document}
%
\title{Tuning Streamed Applications on Intel Xeon Phi: \\A Machine Learning Based Approach}
\author{\IEEEauthorblockN{Peng Zhang\IEEEauthorrefmark{1},
Jianbin Fang\IEEEauthorrefmark{1},
Tao Tang\IEEEauthorrefmark{1},
Canqun Yang\IEEEauthorrefmark{1},
Zheng Wang\IEEEauthorrefmark{2},
}
\IEEEauthorblockA{\IEEEauthorrefmark{1}Compiler Laboratory, College of Computer, National University of Defense Technology,
China \\ Email: \{zhangpeng13a, j.fang, taotang84, canqun\}@nudt.edu.cn}
\IEEEauthorblockA{\IEEEauthorrefmark{2}MetaLab, School of Computing and Communications, Lancaster University, United Kingdom \\
Email: z.wang@lancaster.ac.uk}

}

\maketitle

\begin{abstract}

Many-core accelerators, as represented by the XeonPhi coprocessors and GPGPUs, allow software to exploit spatial and temporal
sharing of computing resources to improve the overall system performance. To unlock this performance potential requires software to
effectively partition the hardware resource to maximize the overlap between host-device communication and accelerator computation, and to
match the granularity of task parallelism to the resource partition. However, determining the right resource partition and task parallelism
on a per program, per dataset basis is challenging. This is because the number of possible solutions is huge, and the benefit of choosing
the right solution may be large, but mistakes can seriously hurt the performance. In this paper, we present an automatic approach to
determine the hardware resource partition and the task granularity for any given application, targeting the Intel XeonPhi architecture.
Instead of hand-crafting the heuristic for which the process will have to repeat for each hardware generation,
we employ machine learning techniques to automatically learn it. We achieve this by first learning
a predictive model offline using training programs; we then use the learned model to predict the resource partition and task granularity
for any unseen programs at runtime. We apply our approach to 23 representative parallel applications and evaluate it on a CPU-XeonPhi mixed
heterogenous many-core platform. Our approach achieves, on average, a 1.6x (upto 5.6x) speedup, which translates to 94.5\% of the performance
delivered by a theoretically perfect predictor.

\end{abstract}


%
\IEEEpeerreviewmaketitle

\section{Introduction}
Heterogeneous many-core systems are now commonplace~\cite{citeulike:2767438}. The combination of using a host CPU
together with specialized processing units (e.g., GPGPUs or the Intel XeonPhi) has been shown in many cases to achieve
orders of magnitude performance improvement. Typically, the host CPU of a heterogeneous platform manages the execution
context while the computation is offloaded to the accelerator or coprocessor. Effectively leveraging such platforms not
only enables the achievement of high performance, but also increases the energy efficiency.

While the heterogeneous many-core design offers the potential for energy-efficient, high-performance computing, software developers are
finding it increasingly hard to deal with the complexity of these systems~\cite{citeulike:14070672}. In particular, programmers need to
effectively manage the host-device communication, because the communication overhead can completely eclipse the benefit of computation off-loading
if not careful~\cite{citeulike:13920330, citeulike:6102210, citeulike:13920339}. \textit{Heterogeneous streaming} has been proposed as a
solution to amortize the host-device communication cost~\cite{DBLP:conf/ipps/NewburnBWCPDSBL16}. It works by partitioning the processor
cores to allow independent communication and computation tasks (i.e. streams) to run concurrently on different hardware resources, which
effectively overlaps the kernel execution with data movements. Representative heterogeneous streaming implementations include CUDA
Streams~\cite{tr:cuda:best}, OpenCL Command Queues~\cite{website:opencl_ref}, and Intel's hStreams~\cite{tr:hstreams:arch,
DBLP:conf/ipps/NewburnBWCPDSBL16}. These implementations allow the program to spawn more than one stream/pipeline so that the data movement
stage of one pipeline overlaps the kernel execution stage of another.

Prior work on heterogeneous streams mainly targets GPUs~\cite{citeulike:13920334, citeulike:9715521, citeulike:13920353}. While also
offering heterogeneous stream execution, the OS-enabled  Intel XeonPhi coprocessor provides some unique features that are currently
unavailable on the GPU. For example, beside specifying the number of streams , developers can explicitly map streams to different groups of
cores on XeonPhi to control the number of cores of each hardware partition. This parameter is not
exposed to programmers on GPUs, making previous work on GPU-based stream optimizations infeasible to fully exploit XeonPhi (see also Section~\ref{sec:alt}). One
the other hand, there are ample evidences showing that choosing the right stream configuration, i.e., the
number of processor core partitions and the number of concurrent tasks of a streamed application,
values, has a significant impact on the streamed application's performance on XeonPhi~\cite{DBLP:conf/npc/LiFTCY16,
DBLP:journals/ppl/FangZLTCCY16}.  However, attempting to find the optimum values through exhaustive search would be ineffective, because
the range of the possible values for the two parameters is huge. What we need is a technique that automatically determines the optimal
stream configuration
for any streamed application in a fast manner.

This paper presents a novel runtime approach to determine the right number of partitions and tasks for heterogeneous streams, targeting the
Intel XeonPhi architecture. We do so by employing machine learning techniques to automatically construct a predictive model to decide at
runtime the optimal stream configuration for
any streamed application. Our predictor is first trained \emph{off-line}. Then, using code and dynamic runtime features of the program,
the model predicts the best configuration for a \emph{new}, \emph{unseen} program. Our approach avoids the pitfalls of using a hard-wired
heuristic that requires human modification every time when the architecture evolves, where the number and the type of cores are likely to change from one generation to the next.

We apply our approach to 23 representative benchmarks, and evaluate it on a heterogeneous many-core platform that has a general purposed
multi-core CPU and a 57-core Intel XeonPhi coprocessor. Our approach achieves, on average, a 1.6x speedup over the optimized, non-streamed
code. This translates to 94.5\% of the best available performance.

We make the following technical contributions:

\begin{itemize}
\item We present the first machine learning model for automatically determining the optimal stream configuration on Intel XeonPhi. Note
that we do not seek to advance the machine learning algorithm itself; instead, we show how machine learning can be used to address the
challenging problem of tuning stream configurations;

\item We develop a fully automatic approach for feature selection and training data generation;
\item We show that our approach delivers constantly better performance over the single-streamed execution across
programs and inputs;
\item Our approach is immediately deployable  and does not require any modification to the application source code.
\end{itemize}

\section{Background and Overview} \label{sec_mlstream_motivation}
In this section, we first give a brief introduction of heterogeneous streams; we then define the scope of this work, before motivating the
need of our scheme and providing an overview of our approach.

\subsection{Heterogeneous Streams} \label{subsec:multiple:streams}
The idea of heterogeneous streams is to exploit temporal and spatial sharing of the computing resources.

\begin{figure}[!t]
\centering
\includegraphics[width=0.5\textwidth]{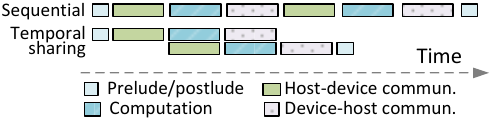}
\caption{Exploit pipeline parallelism through temporal sharing.}
\label{fig:temporal_sharing}
\end{figure}

\cparagraph{Temporal Sharing.} Code written for heterogeneous computing devices typically consists of several stages such as host device
communication and computation. Using temporal sharing, one can overlap some of these stages to exploit pipeline parallelism to improve
performance. This paradigm is illustrated in Figure~\ref{fig:temporal_sharing}. In this example, we can exploit temporal sharing to overlap
the host-device communication and computation stages to achieve better runtime when compared to  execute every stage sequentially.
One way of exploiting temporal sharing is to divide an application into independent tasks so that they can run in a pipeline fashion.

\cparagraph{Spatial Sharing.} Modern many-core accelerators offer a large number of processing units. Since many applications cannot fully
utilize all the cores at a time, we can partition the computing units into multiple groups to concurrently execute multiple tasks. In this
way, the computing resource is spatially shared across concurrently-running application tasks. The key to spatial sharing is to determine
the right number of partitions, because over-provisioning of processing units would waste computing resources but under-provisioning would
lead to slowed down performance.

\lstset{}
\begin{figure}
	\centering %
		\noindent\mbox{\parbox{\columnwidth}{%
\begin{lstlisting}[label=subfig:source_in]
//setting the partition-size and task granularity
hStreams_app_init(partition_size,streams_p_part);

//stream queue id
stream_id = 0;
for(...){
  //enquque host-device transfer to current stream
  hStreams_app_xfer_memory(,,, stream_id, HSTR_SRC_TO_SINK,...);
  ...
  //enqueue computation to the current stream
  hStreams_EnqueueCompute(stream_id, "kernel1", ...);
  ...
  //move to the next stream
  stream_id = (stream_id++) % MAX_STR;
}
//transfer data back to host
hStreams_app_xfer_memory(,,, HSTR_SINK_TO_SRC,...);
\end{lstlisting}%
		}}
    \vspace{-3mm}
	\caption{Example \texttt{hStreams} code.}%
	\label{fig:example_code}%
\end{figure}

\begin{figure*}[!t]
\centering
\subfigure[\texttt{binomial}]{\label{fig_motivation_partition_cf}\includegraphics[width=0.48\textwidth]{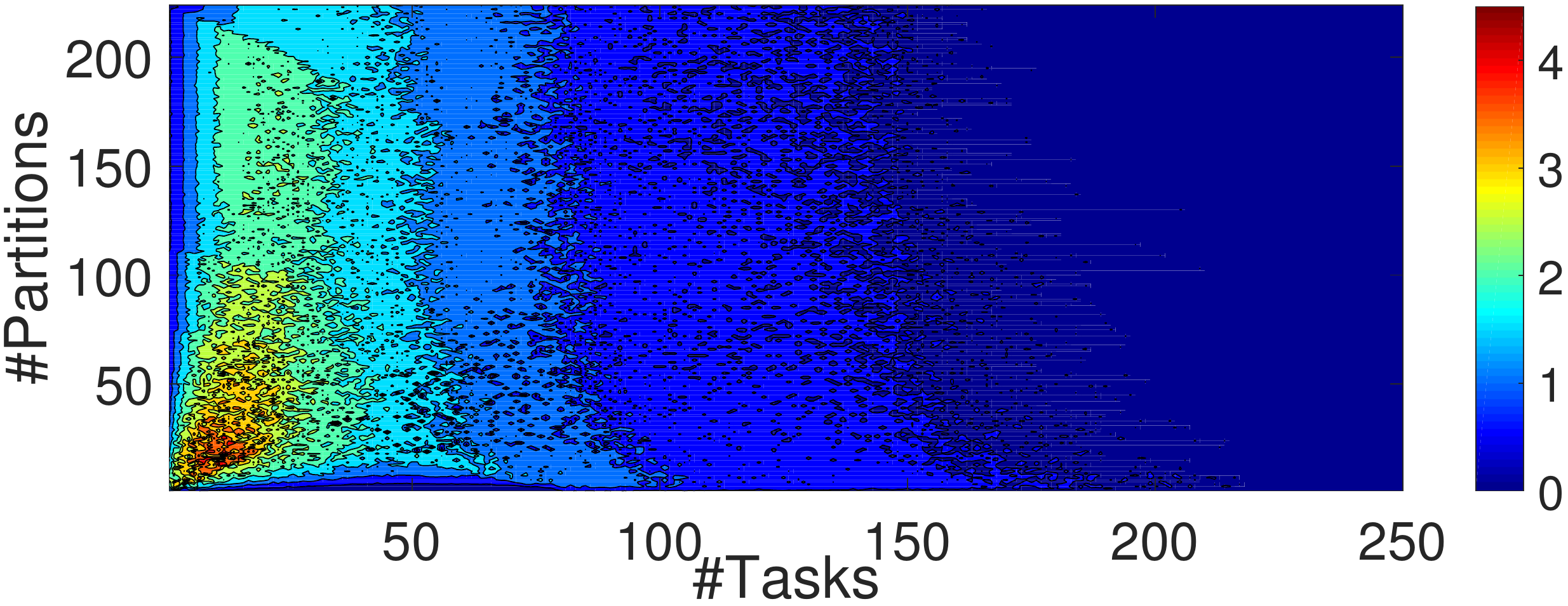}}
\subfigure[\texttt{prefixsum}]{\label{fig_motivation_partition_mm}\includegraphics[width=0.48\textwidth]{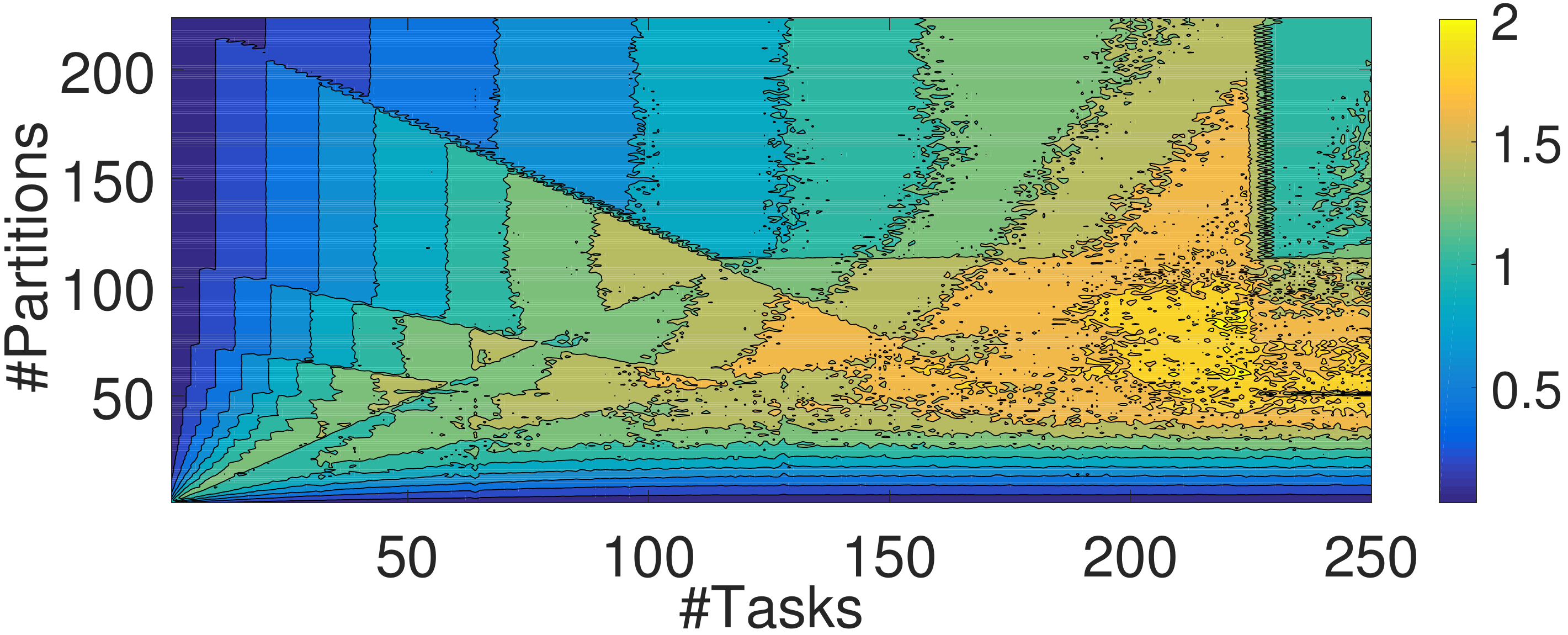}}
\caption{Heatmaps show the resultant speedup (over non-streamed) of \texttt{binomial} and \texttt{prefixsum} under different
stream configurations. The \textit{\#partitions} and \textit{\#tasks}  have a significant impact on the resultant performance, and the sweat spots
are sparse and vary across programs. }
\label{fig_motivation_overall}
\end{figure*}

\subsection{Problem Scope}
Our work aims to improve the performance of a data parallel application by exploiting spatial and temporal sharing of heterogeneous
streams. We do so by determining at runtime how many partitions should be used to group the cores (\emph{\#partitions}) and how many data
parallel tasks (\emph{\#tasks}) should be used to run the application. We target the Intel XeonPhi architecture, but our methodology is
generally applicable and can be extended to other architectures including GPGPUs and FPGAs.

\vspace{1.5mm} \noindent \textbf{Code Example.} Figure~\ref{fig:example_code} gives a simplified code example written with Intel's
\texttt{hStreams} APIs. At line 2 we initialize the stream execution by setting the number of partitions and tasks/streams per partition.
This initialization process essentially creates multiple processor domains and determines how many logical streams can run on a partition.
In the \emph{for} loop (lines 7-14) we enqueue the communication and computation tasks to a number of streams identified by the
\texttt{stream\_id} variable. In this way, communication and computation of different streams can be overlapped during execution (temporal
sharing); and streams on different processor domains (or partitions) can run concurrently (spatial sharing). Our predictive model
determines the \textit{\#partitions} and the \textit{\#tasks} before invoking the \texttt{hStreams} initialization routine,
\texttt{hStreams\_app\_init()}. We also create a wrap for this API to automatically invoke our predictive model, so no modification to the
application source code is required.

\begin{figure}
  \centering
  \includegraphics[width=0.5\textwidth]{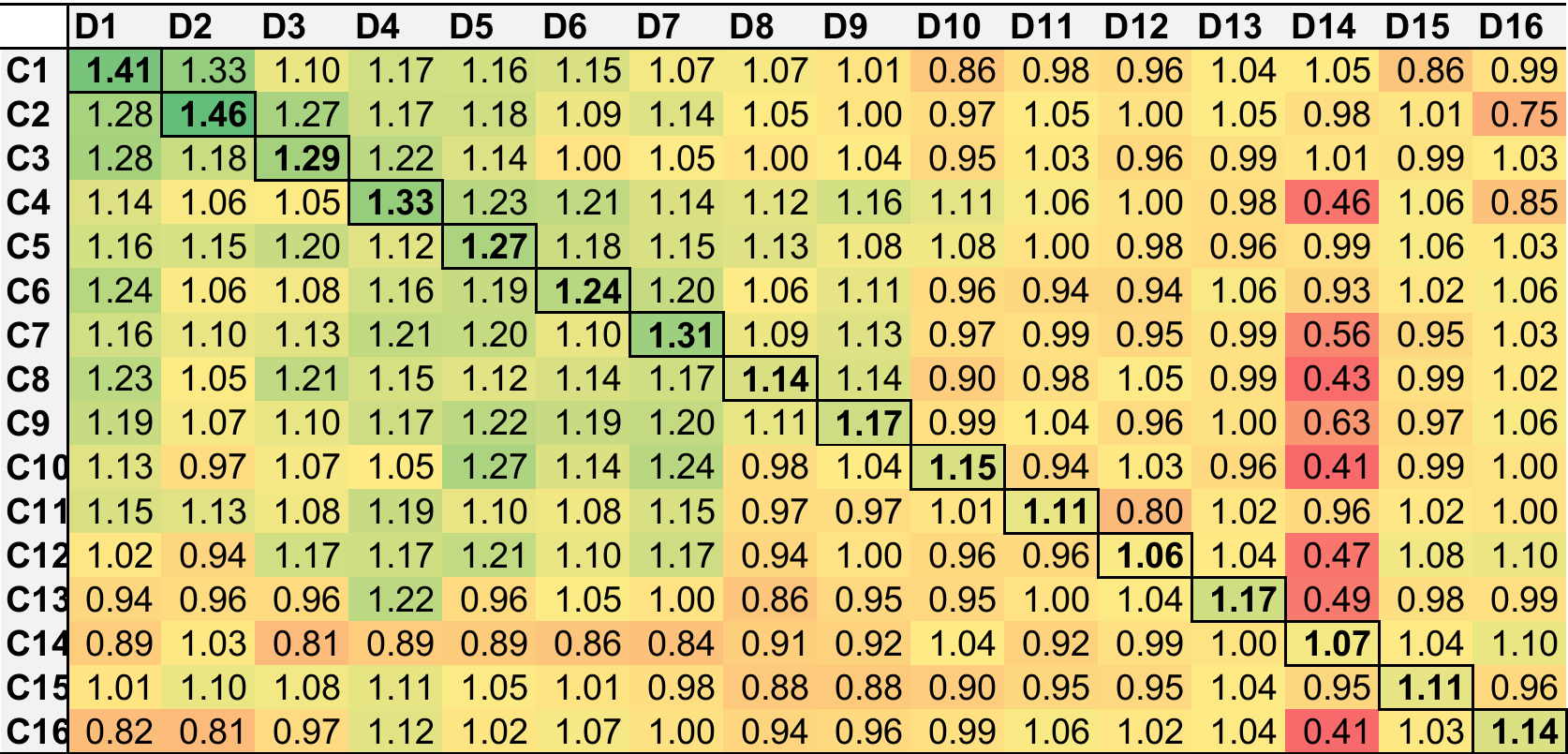}\\
  \caption{Colour table showing the speedups of best-performing configurations across inputs for \texttt{dct}. Each cell shows
  the performance for one of the 16 best-performing configurations, $Cn$, on a given input, $Dn$.
  The best configuration varies across inputs and a good configuration on one input
  can give poor performance on another dataset.}\label{fig:acrossdatasets}
\end{figure}

\subsection{Motivating Examples} \label{subsec:motivate:example}

Consider Figure~\ref {fig_motivation_overall} which shows the resultant performance improvement over the non-streamed version of the code
for two applications on a 57-core Intel XeonPhi system. It is observed from this example that no all stream configurations give  improved
performance. As can be seen from the diagrams, the search space of stream configuration is huge but good configurations are sparse. The
performance varies significantly over stream configurations (\textit{\#partitions}, \textit{\#tasks}). The optimal \textit{\#tasks} for
\texttt{binomial} ranges from 1 to 30, and the best \textit{\#partitions} is between  1 and 40. In contrast to \texttt{binomial},
\texttt{prefixsum} benefits from fine-grained parallelism when using a larger \textit{\#tasks} (220 to 224) and \textit{\#partitions} (60
to 80). However, the stream configurations that are effective for \texttt{prefixsum} give no speedup over the non-streamed version for
\texttt{binomial}.

Now consider Figure~\ref{fig:acrossdatasets} that shows the speedups of \texttt{dct} under 16 configurations over the non-streamed version,
where each configuration is found to give the best-performance for one of the 16 inputs. In the color table, each cell shows the
performance of a stream configuration ($C1, ..., C16$) on a specific input dataset ($D1, ..., D16$); and the values along the diagonal line
represent the best-available performance for an input. As can be seen from the figure, the best stream configuration can vary across inputs
for the same benchmark. For example, while $C4$ gives 1.33 speedup for dataset $D4$, it delivers a poor performance for dataset $D14$ by
doubling the execution time over the non-streamed version. This diagram also suggests that none of the 16 configurations gives improved
performance for all inputs.


\cparagraph{Lesson Learned.} These two examples demonstrate that choosing the stream configuration has a great impact on the resultant
performance and the best configuration must be determined on a per-program and per-dataset basis.  Attempting to find the optimal
configuration through means of an exhaustive search would be ineffective, the overhead involved would be far bigger than the potential
benefits. Online search algorithms, while can speedup the search process, the overhead can still outweigh the benefit. For example, when
applying simulated annealing to \texttt{binomial}, the best-found configuration only reaches 84\% of the best-available performance after
310,728 iterations\footnote{Later in Section~\ref{sec:overall}, we show that our approach achieves 98\% of the best-available performance
for \texttt{dct}.}. Classical hand-written heuristics are not ideal either, as they are not only complex to develop, but are likely to fail
due to the variety of programs and the ever-changing hardware architecture. An alternate approach, and the one we chose to use, is to use
machine learning to automatically construct a predictive model directly predict the best configuration, providing minimal runtime, and
having little development overhead when targeting new architectures.

\subsection{Overview}
\begin{figure}
  \centering
  \includegraphics[width=0.5\textwidth]{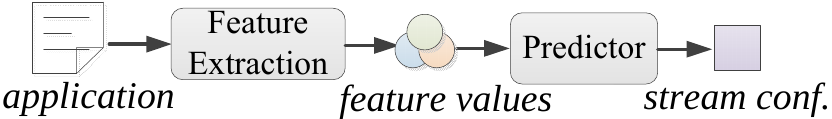}\\
  \caption{Our machine learning based  model
  predicts the optimal stream configuration based on the code and runtime features.
  }\label{fig:workflow}
\end{figure}

Our library-based approach, depicted in Figure~\ref{fig:workflow}, is completely automated. To determine the best streaming configuration,
our approach follows a number of steps described as follows. We use a set of information or \emph{features} to capture the characteristics
of the program. We develop a LLVM~\cite{Lattner:2004:LCF:977395.977673} compiler pass to extract static code features at compile time, and
a low-overhead profiling pass to collect runtime information at execution time. Because profiling also contributes to the final program
output, no computation cycle is wasted. At runtime, a predictive model (that is trained offline) takes in  the feature values and predicts
the optimal stream configuration. The overhead of runtime feature collection and prediction is a few milliseconds, which is included in all
our experimental results.

\section{Predictive Modeling} \label{sec_mlstream_modeling}

Our model for predicting the best stream configuration is a Support Vector Machine (\SVM) with a quadratic function kernel. The model is
implemented using libSVM (C++ version)~\cite{libsvm}. We have evaluated a number of alternative modeling techniques, including regression,
K-Nearest neighbour (\KNN), decision trees, and the artificial neural network (\ANN), etc. We chose \SVM because it gives the best
performance and can model both linear and non-linear problems (Section~\ref{sec_compare_learning_techniques}). The model takes in feature
values and produces a label for the optimal stream configuration.

Building and using such a model follows a 3-step process for supervised learning: (i) generate training data (ii) train a predictive model
(iii) use the predictor, described as follows.

\begin{figure}
  \centering
  \includegraphics[width=0.5\textwidth]{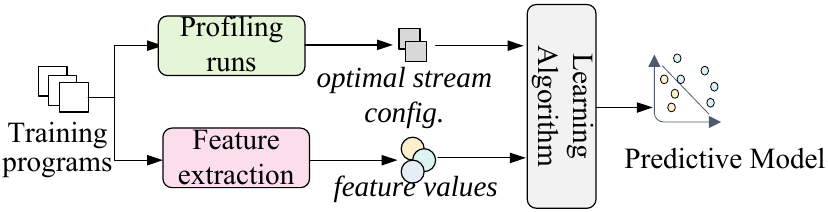}\\
  \caption{The training process of our approach}\label{fig:training}
\end{figure}

\subsection{Training the Predictor \label{sec:training}}
Our method for model training is shown in Figure~\ref{fig:training}. To learn a new predictor we first need to find the best stream
configuration for each training program, and extract the feature values from the program. We then use this set of feature values and
optimal configurations to train a model.

\subsubsection{Generating Training Data} We use cross validation by excluding the testing benchmarks from the training dataset. To generate
the training data for our model we used 15 programs. We execute each training program and benchmark a number of times until the gap of the
upper and lower confidence bounds is smaller than 5\% under a 95\% confidence interval setting. We exhaustively execute each training
program across all of our considered stream configurations, and record the performance of each. Specifically, we profile the program using
the \emph{\#partitions} ranging from 1 to 224 and the \emph{\#tasks}  ranging from 1 to 256~\footnote{We chose these values because
configuration settings beyond these values give poor performance during our initial evaluation.}. Next, we record the best performing
configuration for each program and dataset, keeping a label of each. Finally, we extract the values of our selected set of features from
each program and dataset.

\vspace{1.5mm} \noindent \textbf{Data Labeling.} 
Our initial labeling process generates over 100 labels while we only have a small number of training samples. The expected predicting
accuracy is deemed to be low as some of the labels only have a handful of examples. Therefore, we have to merge labels after generating the
raw training data. Our label merging procedure is shown in Algorithm~\ref{alg_label_merging}. This merging process aims to reduce the
number of labels to an order of magnitude less than that of samples (from $N$ to a configurable parameter $Nr$). The input are the training
samples, each with a set of well-performing stream configurations (e.g., the top 3\% best-performing configurations). We calculate the
similarity of two labels using three quantitative metrics: (a) the common best configurations (we aim to keep the common best
configurations), (b) whether the samples are from the same program, and (c) whether they are with the same dataset. The three metrics are
denoted by $\omega_1$, $\omega_2$, and $\omega_3$ respectively in Algorithm~\ref{alg_label_merging}. We discuss the performance impact of
the label merging algorithm in Section~\ref{sec:dist}. Then, we sort the weights in a descending order, merge corresponding labels, and
update the label for each sample. The output samples are labeled with merged classes. Applying the data labeling process described above
results in 28 labels (i.e., $Nr$=28).

\begin{algorithm}[t!]
\caption{The Label Merging Algorithm.}\label{alg_label_merging}
\begin{algorithmic}[1]
\scriptsize
{ \LState {\textbf{input:} LS -- label set, PS -- program set, DS -- dataset} \LState {\textbf{output:} LSr -- reduced label
set} \Procedure {Merging\_Labels}{$LS$, $PS$, $DS$, $N$; $LSr$, $Nr$}
	\LState $LSr \leftarrow LS, Nr \leftarrow N$
	\Repeat
		\For{$s \leftarrow 0, Nr$}
			\For{$t \leftarrow 0, Nr$}
				\LState $p_{s} \leftarrow PID(PS, s), p_{t} \leftarrow PID(PS, t)$
				\LState $d_{s} \leftarrow DID(DS, s), d_{t} \leftarrow DID(DS, t)$
				\If {$s \neq t$}
					\LState $w_{1}, w_{2}, w_{3} \leftarrow 0$
					\LState $LS_{st}\leftarrow LS_{s} \sqcap LS_{t}$
					\LState $w_{1} \leftarrow Counts(LS_{st})$
					\If {$p_{s} == p_{t}$} \Comment{the same program}
						\LState $w_{2} \leftarrow 150$
					\EndIf
					\If {$d_{s} == d_{t}$} \Comment{the same dataset}
						\LState $w_{3} \leftarrow 30$
					\EndIf
					\LState $w(s,t) \leftarrow w_{1}+w_{2}+w_{3}$
				\EndIf
			\EndFor
		\EndFor
		\LState $RankOnWeight(w)$ \Comment{Rank w on the weights}
		\LState $MergeAndUpdate(w, LS, N, LSr, Nr)$
	\Until{$LS_{i} \sqcap LS_{j}=\phi, \forall i,j\in Nr  \wedge i \neq j$}	
\EndProcedure
}
\end{algorithmic}
\end{algorithm}

\subsubsection{Building The Model}
The corresponding configuration labels, along with the feature values for all training programs, are passed to a learning algorithm. The
 algorithm finds a correlation between the feature values and the optimal stream configuration. The output of our
learning algorithm is a \SVM model where the weights of the model are determined from the training data. We use the parameter tuning tool
provided by libSVM to determine the kernel parameters. Parameter search is performed on the training dataset using cross-validation. In our
case, the overall training process (which is dominated by training data generation) takes less than a week on a single machine. Since
training is performed only once ``at the factory'', this is a \emph{one-off} cost.

\subsection{Features} \label{sec_mlstream_modeling_features}
Our predictive models are based exclusively on code and dynamic features of the target programs. Code features are extracted from the
program source code, and dynamic features are collected using hardware performance counters during the initial profiling run of the target
application.  We restrict us to use hardware performance counters that are commonly available on modern processors such as the data cache
misses to ensure that our approach can be applied to a wide range of architectures.

We considered 38 candidate raw features in this work. Some features were chosen from our intuition based on factors that can affect the
performance such as \texttt{dts} (host-device data transfer size) and \texttt{\#xfer\_mem}, while other features were chosen
based on previous work~\cite{fursin2008milepost,DBLP:journals/taco/WangGO14}.

\subsubsection{Feature Selection} \label{sec_feature_selection}
To build an accurate predictive model through supervised learning, the training sample size typically needs to be at least one order of
magnitude greater than the number of features. In this work, we start from 280 training  samples and 38 raw features, so we would like to
reduce the number of features in use. Our process for feature selection is fully automatic, described as follows. We first combine several
raw features to form a set of combined normalized features, which are able to carry more information than the individual parts. For
example, instead of reporting raw branch hit and miss counts we use the branch miss rate. Next, we removed raw features that carried
similar information which is already captured by chosen features. To find which features are closely correlated we constructed a
correlation coefficient matrix using the Pearson correlation coefficient. The closer a coefficient between two features is to +/-1, the
stronger the correlation between the two input features. We removed any feature which had a correlation coefficient (taking the absolute
value) greater than 0.7. Similar features include the number of executed instructions and the number of E-stage cycles that were
successfully completed.  Our feature selection process reduces the number of features to 10, which are listed in Table~\ref{tbl_features}.
Since our approach for feature selection is automatic, the approach can be applied to other sets of candidate features. It is to note that
feature selection is also performed using cross-validation (see also Section~\ref{sec:compa}).

\begin{table}[!t]
\scriptsize
\caption{Final selected features.}
\vspace{-4mm}
\begin{center}
\begin{tabular}{ll}
\toprule
\textbf{Feature} & \textbf{Description} \\
\midrule
\rowcolor{Gray}  loop nest & at which level the loop can be parallelized\\
loop count & \# of the parallel loop iterations \\
\rowcolor{Gray}  \#xfer\_mem & \# of host-device transfer API calls\\
dts & total host-device transfer size\\
\rowcolor{Gray}  redundant transfer size & host-device transfer size among overlapping tasks\\
max blocks & the maximum number of tasks of the appliation\\
\rowcolor{Gray}  min task unit &  the minimum task granularity for a partition\\
\# instructions & the total number of instructions of the kernel\\
\rowcolor{Gray}  branch miss & branch miss rate\\
L1 DCR & L1 Data cache miss rate\\
\bottomrule
\end{tabular}
\end{center}
\label{tbl_features}
\end{table}

\subsubsection{Feature Scaling}
Supervised learning typically works better if the feature values lie in a certain range. Therefore, we scaled the value for each of our
features between the range of 0 and 1. We record the maximum and minimum value of each feature found at the training phase, and use these
values to scale features extracted from a new application after deployment.

\begin{figure}
  \centering
  \includegraphics[width=0.5\textwidth]{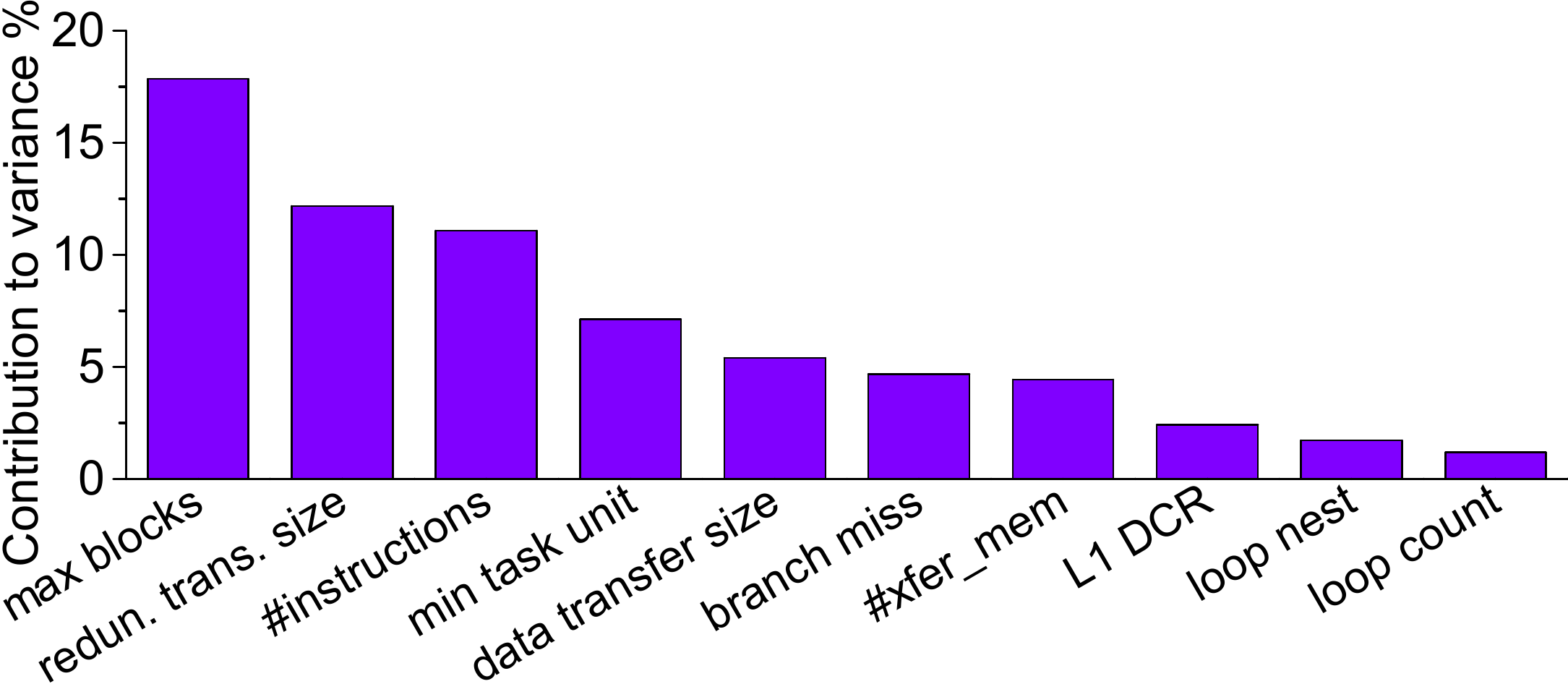}
  \caption{Feature importance according to the Varimax rotation.}\label{fig:pca}
\end{figure}

\subsubsection{Feature Importance}
To understand the usefulness of each selected feature,
we apply a factor analysis technique called Varimax
rotation~\cite{manly2004multivariate} to the feature space transformed
by the principal component analysis (\PCA).
 This technique
quantifies the contribution of each feature to the overall variance in each
of the \PCA dimensions. Intuitively, the more variances a feature brings
to the space, the more useful information the feature carries. Features that
capture the parallelism degree (e.g. \texttt{max blocks}), host-device
communication (e.g. \texttt{redundant transfer size}), and computation (e.g.
\texttt{\#instructions}) are found to be important. Other features such as
\texttt{L1 DCR} and \texttt{loop nest} are useful, but are less important
compared to others. This figure
shows that prediction can accurately draw upon a subset of
aggregated feature values.

\subsection{Runtime Deployment}
\vspace{-1mm}
 Once we have built and trained our predicted model as described above, we can use it to predict the best stream configuration
for any \emph{new}, \emph{unseen} program. When an application is launched, we will first extract the feature values of the program. Code
features (such as \texttt{loop count}) are extracted from the program source. Dynamic features (such as \texttt{branch miss}) are extracted
by profiling the program without partitioning for several microseconds. After feature collection, we feed the feature values to the offline
trained model which outputs a label indicating the stream configuration to use for the target program.

\cparagraph{Adapt to Changing Program Phases.} Our approach can adapt to different behaviors across kernels because predictions are
performed on a per-kernel basis. It can be extended to adapt phase changes within a kernel. This can be achieved by checking periodically
sampling if the performance counter readings are significantly different from the ones use for the initial prediction to trigger
re-prediction and re-configuration. Dynamic re-configuration will require extending \texttt{hStreams} to adjust thread mapping and having
hardware support to stop and resume the execution contexts.

\section{Experimental Setup} \label{sec_mlstream_setup}

\subsection{Hardware, System Software and Benchmarks} \label{subsec:benchmarks}
\vspace{-2mm}
 \cparagraph{Platform.} Our evaluation platform is an Intel Xeon server with an Intel dual-socket 8-core Xeon CPU @ 2.6 Ghz
(16 cores in total) and an Intel Xeon 31SP Phi accelerator (57 cores). The host CPUs and the accelerator are connected through
\texttt{PCIe}. The host environment runs Redhat Linux v7.0 (with kernel v.3.10). The coprocessor environment runs a customized uOS
(v2.6.38.8). We use Intel's MPSS (v3.6) to communicate between the host and the coprocessor and Intel's ~\texttt{hStreams} library (v3.6).

\cparagraph{Benchmarks.} As currently there exist very few programs written with Intel's \texttt{hStreams}, we faithfully translated 21
applications to \texttt{hStreams} from the commonly used benchmark suites\footnote{Our benchmarks can be downloaded from
\url{https://github.com/Wisdom-moon/hStreams-benchmark.git}.}. Table~\ref{tbl_benchmarks} gives the full list of these benchmarks.
Among them, \texttt{convolutionFFT2d} and \texttt{convolutionSeparable} have algorithm-dependent parameters, which are regarded as
different benchmarks in the experiments. This setting gives us a total of 23 programs. We run the majority of the programs using over 25
different datasets, except for some applications where we used around 10 datasets because the algorithmic constraints of the applications
prevent us from generating a large number of inputs.


\begin{table}[!t]
\scriptsize
\caption{Programs used in our experiments.}
\vspace{-5mm}
\begin{center}
\begin{tabular}{lllll}
\toprule

\textbf{Suite} & \textbf{Name} & \textbf{Acronym} & \textbf{Name} & \textbf{Acronym}\\
\midrule

\rowcolor{Gray}         & convol.Separable  & convsepr1(8) & dotProduct &  dotprod  \\
\rowcolor{Gray}         & convolutionFFT2d  & fftx1y1(4y3) & fwt &  fwt  \\
\rowcolor{Gray}         &  MonteCarlo & montecarlo   & matVecMul  & mvmult  \\
\rowcolor{Gray}         & scalarProd &  scalarprod &  transpose  & transpose \\
\rowcolor{Gray}         \multirow{-6}{0.01\textwidth}{NVIDIA SDK}
                        & vectorAdd &  vecadd  & &\\

\multirow{2}{0.01\textwidth}{AMD SDK}
       & binomial & binomial  & BlackScholes & blackscholes \\
       & dct  & dct & prefixSum & prefix  \\

\rowcolor{Gray}  	& bfs & bfs  & histo&  histo  \\
\rowcolor{Gray}         & lbm & lbm & mri-q &  mri-q  \\
\rowcolor{Gray}         & mri-gridding & mri-gridding &  sad & sad \\
\rowcolor{Gray}  	\multirow{-3}{*}{Parboil}
			& sgemm &  sgemm & spmv  & spmv \\
\bottomrule
\end{tabular}
\end{center}
\label{tbl_benchmarks} 
\end{table}

\subsection{Competitive Approaches\label{sec:compa}}
Because there is currently no expert-tuned heuristic for choosing stream configurations on XeonPhi, we compare our approach against two
recent models for predicting the optimal stream configuration on GPUs. As it is currently not possible to configure the number of
partitions  on GPUs, the relevant models can only predict the number of tasks (or streams).

\subsubsection{\Modela} In ~\cite{citeulike:13920353}, Liu \etal use linear regression models to search for the optimal number of tasks for GPU programs~\cite{citeulike:13920353}.
The approach employs several analytic models described as follows.

For a task with an input data size of $m$, the transferring time  between the CPU and the GPU, $T_t$, is determined as $T_t = \alpha \cdot
m + \beta$, and the computation time, $T_c$, is calculated as: $T_c=\eta \cdot m+\gamma$ where the model coefficients, $\alpha$, $\beta$,
$\eta$ and $\gamma$, are determined through empirical experiments.
For a given kernel with $N$ input data elements running using $n$ streams, this approach partitions the computation into $n$ tasks, where
the data size for each task, $m$, is equal to $N$/$n$. Therefore, the total execution time, $T_{total}$, can be determined by:
$$T_{total} =T_{t} + nT_{c}=\alpha \cdot m+\frac{N\gamma}{m}+N\eta+\beta$$ By calculating the partial differential and second-order partial differential of
$T_{total}$ with respect to $m$, we can obtain the optimal task-granularity as $m= \sqrt{\frac{N\gamma}{\alpha}}$.
Then we can calculate the number of tasks ($n$). Note that, we set the \textit{\#partitions} to be the same as $n$ for XeonPhi.

\subsubsection{\Modelb} The work presented by Werkhoven \etal models the performance of data transfers between
the CPU and the GPU~\cite{citeulike:13920334}. They use the LogGP model to estimate the host-device data transfer time. Specifically,
the model estimates the data transfer time using five parameters: the communication latency ($L$), overhead ($o$), the gap ($g$), the
number of processors ($P$), and the \texttt{PCIe} bandwidth ($G$).

Let $B_{hd}$ denotes the amount of data transferred from the host to the device and $B_{dh}$ denotes vice versa, and $T_{kernel}$ donates
the kernel execution time. Then, the optimal number of streams (i.e., \emph{\#tasks}), $N_s$,  can be estimated by solving the following
equations:
\begin{small}
\[
B_{dh} * G_{dh} + g*(N_s-1) =
\begin{cases}
\frac{T_{kernel}}{N_s}+\frac{B_{dh}}{N_s} *G_{dh}, & \text{if} B_{dh} > B_{hd}\\
\frac{B_{hd}}{N_s}*G_{hd}+\frac{T_{kernel}}{N_s}, & \text{otherwise}
\end{cases}
\]
\end{small}
Again, for this model, we set the \textit{\#partitions} to be equal to the optimal $N_{s}$ value on XeonPhi.

\subsection{Evaluation Methodology}

\noindent \textbf{Model Evaluation.} We use  cross-validation to evaluate our machine learning model. Our model is trained using benchmarks
from the AMD and NVIDIA SDK suites, we then apply the trained model to benchmarks from the Parboil suite. We apply \emph{leave-one-out}
cross validation to the AMD and NVIDIA SDK suites.  This means that we exclude the target program from the training program set, and learn
a model using the remaining programs from the AMD and NVIDIA suites; we then then apply the learnt model to the testing program. We repeat
this process to ensure each benchmark from the AMD and NVIDIA suites is tested. It is a standard evaluation methodology, providing an
estimate of the generalization ability of a machine-learning model in predicting \emph{unseen} data. Note that we exclude both
\texttt{convolutionFFT2d} and \texttt{convolutionSeparable} from the training set when one of the two is evaluated.

\vspace{1.5mm} \noindent \textbf{Performance Report.} We run each program under a stream configuration multiple times and report the
\emph{geometric mean} of the runtime. To determine how many runs are needed, we calculated the confidence range using a 95\% confidence
interval and make sure that  the difference between the upper and lower confidence bounds is smaller than 5\%.


\section{Experimental Results} \label{sec_mlstream_results}
In this section, we first present the overall performance of our approach.
We then compare our approach to the fixed stream configuration and the two competitive approaches and before discussing the working mechanism of our scheme.

\subsection{Overall Performance}
\begin{figure}[!t]
  \centering
  \includegraphics[width=0.5\textwidth]{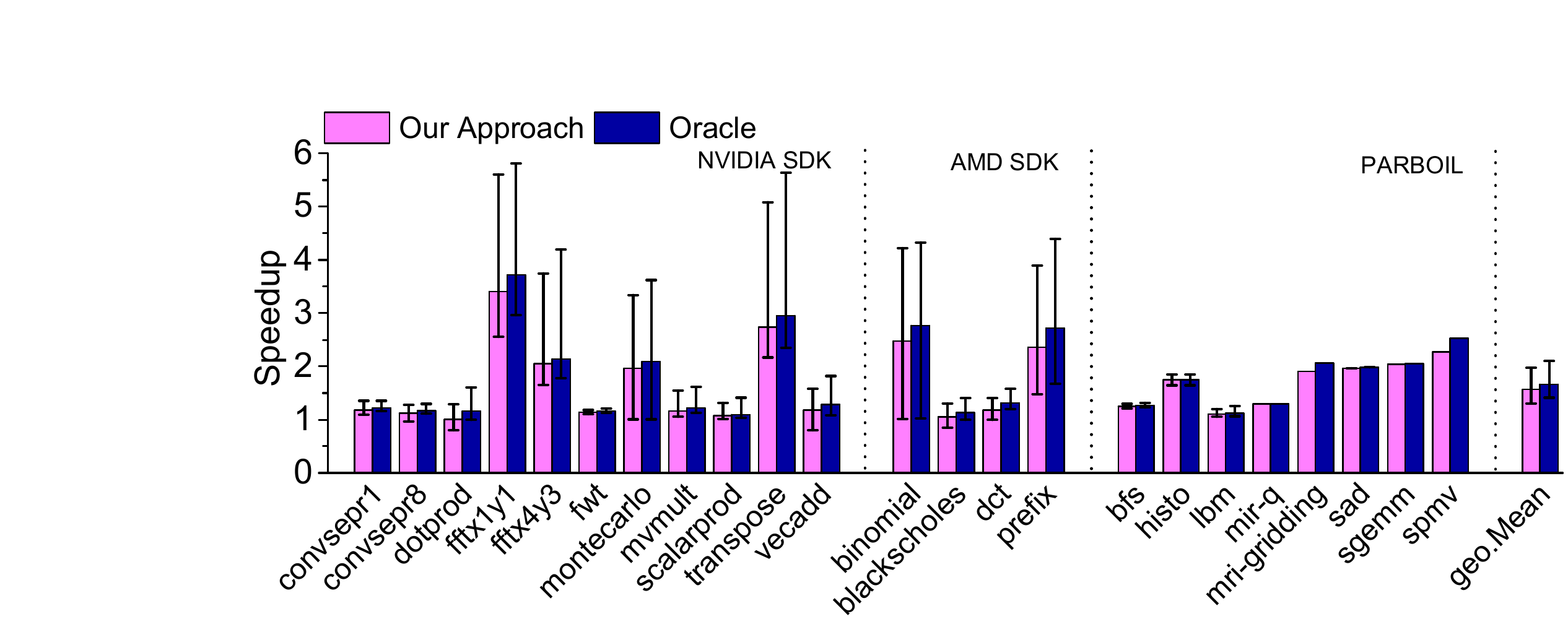}\\
  \caption{Overall performance compared to a single stream version. Our approach achieves, on average, 94.5\% of the oracle performance. The min-max
bars show the range of performance achieved across different inputs. }\label{fig_overall_perf}
\end{figure}

\begin{figure}[t!]
  \centering
  \includegraphics[width=0.5\textwidth]{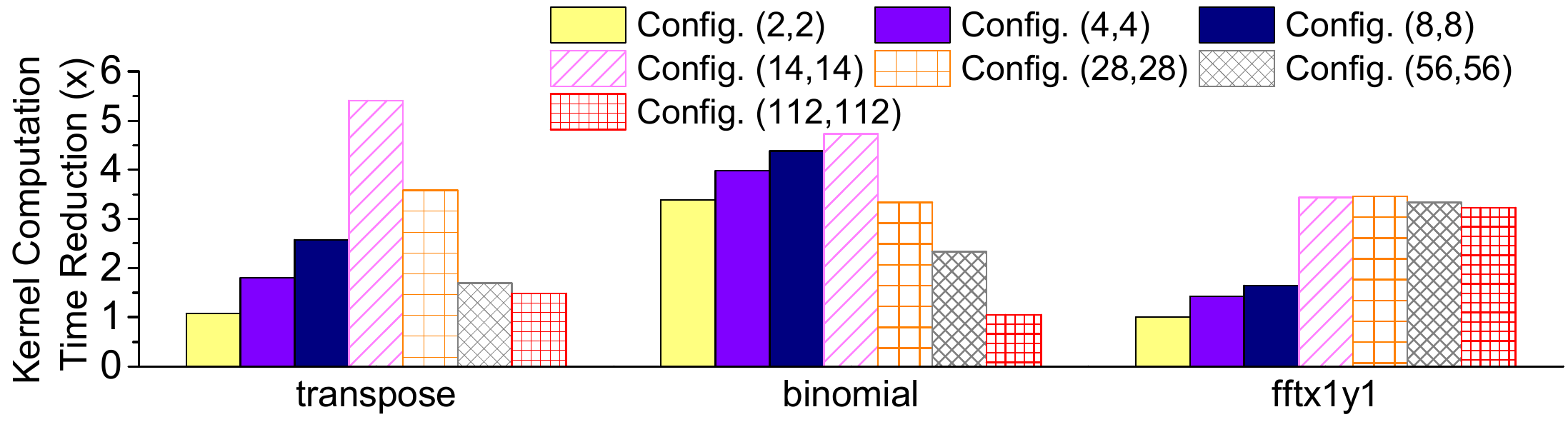}\\
  \caption{Reduction of kernel computation time over a single stream execution. The performance improvement comes from the reduction of the threading overhead.
  A stream configuration is annotated as (\emph{\#partitions}, \emph{\#tasks}). }
  \label{fig:extra_speedup}
\end{figure}

In this experiment, we exhaustively profiled each application with all possible stream configurations and report the best-found performance
as the Oracle performance. The Oracle performance gives an indication of how close our approach is to a theoretically perfect solution. The
baseline used to calculate the speedup is running the application using a single stream without processor core partitioning.

\subsubsection{Overall Results}
\label{sec:overall} The result is shown in Figure~\ref{fig_overall_perf}. The min-max bar on the diagram shows the range of speedups per
application across all evaluated inputs. Overall, our approach achieves an average speedup of 1.6$\times$ over the non-streamed code. This
translates to 94.5\% of the Oracle performance. Although our model is not trained on the Parboil benchmark suite, it achieves good
performance, delivering 97.8\% of the Oracle performance on this benchmark suite. This demonstrates the portability of our approach across
benchmarks.

\subsubsection{Analysis of High Speedup Cases}
\label{sec:hsc}
 We found that there are several benchmarks obtain a speedup of over 2$\times$. After having a closer investigation, we
notice that such performance is because that streaming can also reduce the kernel execution time for these applications.

To quantify the benefit of kernel time reduction, we measure the kernel execution time with and without multiple streams and calculate the
speedup between them. Note that we \emph{exclude the host-device communication time in this case}. The kernel time improvement for
\texttt{transpose}, \texttt{binomial}, and \texttt{fftx1y1} is shown in Figure~\ref{fig:extra_speedup}. As can be seen from the diagram,
choosing a good stream configuration can lead to more than 4x reduction on the kernel execution time. This is because these benchmarks are
implemented by parallelizing the inner loop within a nested loop. During runtime, the parallel threads working on the inner loop will need
to be created, synchronized, or destroyed for each outer loop iteration. This threading overhead could be significant when the outer loop
iterates many times. When using multiple streams, we essentially divide the whole outer loop iteration space into multiple smaller
iteration space. This allows multiple groups of threads to be managed simultaneously, leading to a significant decrease in threading
overhead and faster kernel execution time. On the other hand, we note that using too many streams and partitions will lead to a performance
decrease. This is due to the fact that stream management also comes at a cost, which increases as the  number of partitions increases.
Nonetheless, for applications where the kernel computation domains the program execution time, by reducing the kernel time can lead to
additional improvement, yielding more than 2x speedups.

\subsubsection{Speedup Distribution}
\begin{figure}[t!]
  \centering
  \includegraphics[width=0.5\textwidth]{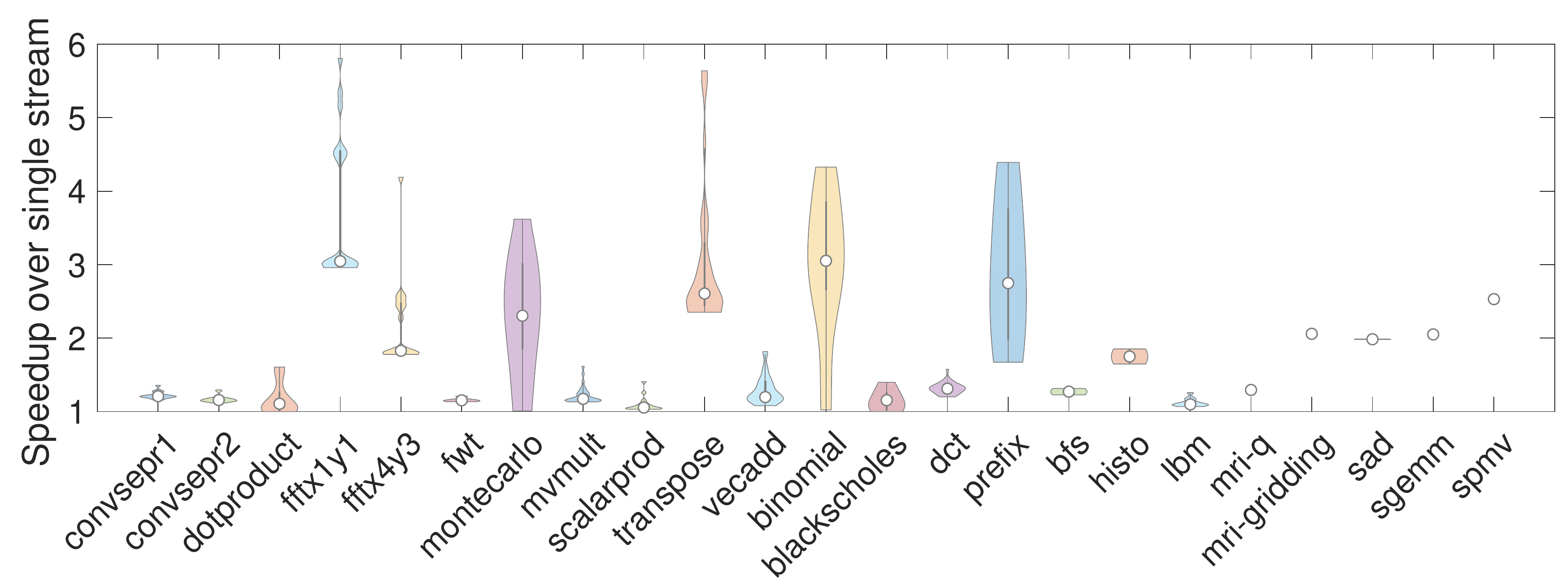}\\
  \vspace{-2mm}
  \caption{Violin plot showing the speedups per benchmark across datasets.
  The shape of the violin corresponds to the speedup distribution. The thick black line shows where 50\% of the
    data lies.
  }
  \label{fig:individualdist}
\end{figure}

We show the speedups per benchmark across datasets in Figure~\ref{fig:individualdist}. The shape of the violin plot corresponds to the
speedup distribution. We see that the speedups of \texttt{montecarlo} and \texttt{prefix} distribute fairly uniformly while the data
distribution of \texttt{fftx1y1} and \texttt{fftx4y3} is multimodal (i.e. it has two peaks). Further, the input datasets have little impact
on the behavior of \texttt{fwt} and \texttt{lbm} so the speedups remain constant across datasets.
To conclude, the streaming speedups of some applications are sensitive to the input datasets while that of others are not.

\begin{figure}[!t]
  \centering
  \includegraphics[width=0.5\textwidth]{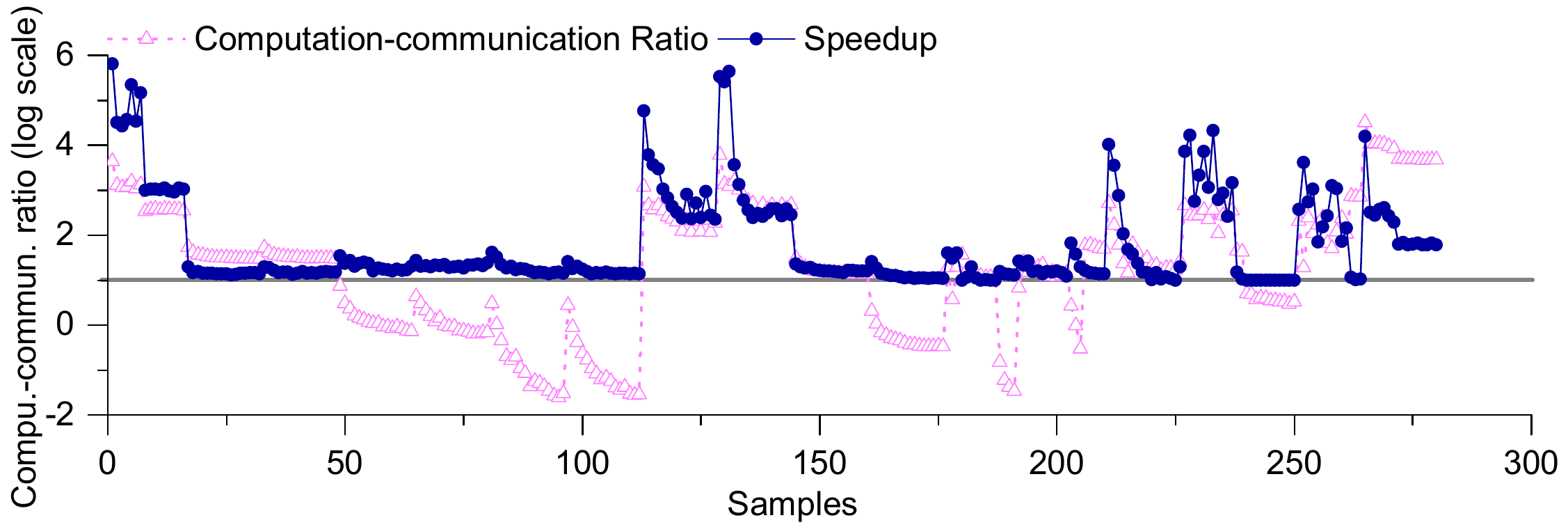}\\
  \vspace{-2mm}
  \caption{The relation between computation-communication ratio and the speedup. The computation-communication ratio is normalized using the natural logarithm function. Thus, the kernel computation time equals the host-device communication time when $ratio=0$.
  In general, a higher computation-communication ratio leads to a better speedup.
  }
  \label{fig_overall_ratio_sp}
\end{figure}
\subsubsection{Correlation Analysis}
Figure~\ref{fig_overall_ratio_sp} shows the relation between the computation-communication ratio and the achieved speedup when using
heterogeneous streams across all benchmarks and datasets. We see that the computation-communication ratio varies over the benchmarks and
the speedup changes accordingly, but in general a higher computation-to-communication ratio leads to a greater speedup. As explained in
Section~\ref{sec:hsc}, in addition to overlapping the computation and communication, our approach can also reduce the kernel computation
time by choosing the right stream configuration. Therefore, benchmarks with a high computation-communication ratio also benefit from a
reduction in the kernel computation time.

 To quantify the
relation between the computation-communication ratio and the speedup, we calculate the Pearson correlation coefficient of the two
variables. The calculation gives a correlation coefficient of 0.7, indicating that the two variables (the computation-communication ratio
and the speedup) have a strong linear correlation. By carefully selecting the stream configuration, our approach tries to maximize the
overlap between communication and computation, which thus leads to favourable performance.

\vspace{1.5mm}
\noindent \textbf{Summary.} The performance improvement of our
approach comes from two factors. First, by predicting the right processor
 partition, our approach allows effective overlapping of the host-device
communication and computation.
Second, by matching task parallelism to the resource partition, our approach
can reduce the overhead of thread management, compared to the single stream execution.
When the host-device communication time dominates
the streaming process, the performance improvement mainly comes from
computation-communication overlapping and the speedup from streaming is consistantly less than 2$\times$.
When the kernel execution time dominates the stream process, the application can benefit from the overhead reduction of thread management.
In this case, the speedup can be as large as 5$\times$.
This trend can be clearly seen from Figure~\ref{fig_overall_ratio_sp}.

\begin{figure}[!t]
  \centering
  \includegraphics[width=0.5\textwidth]{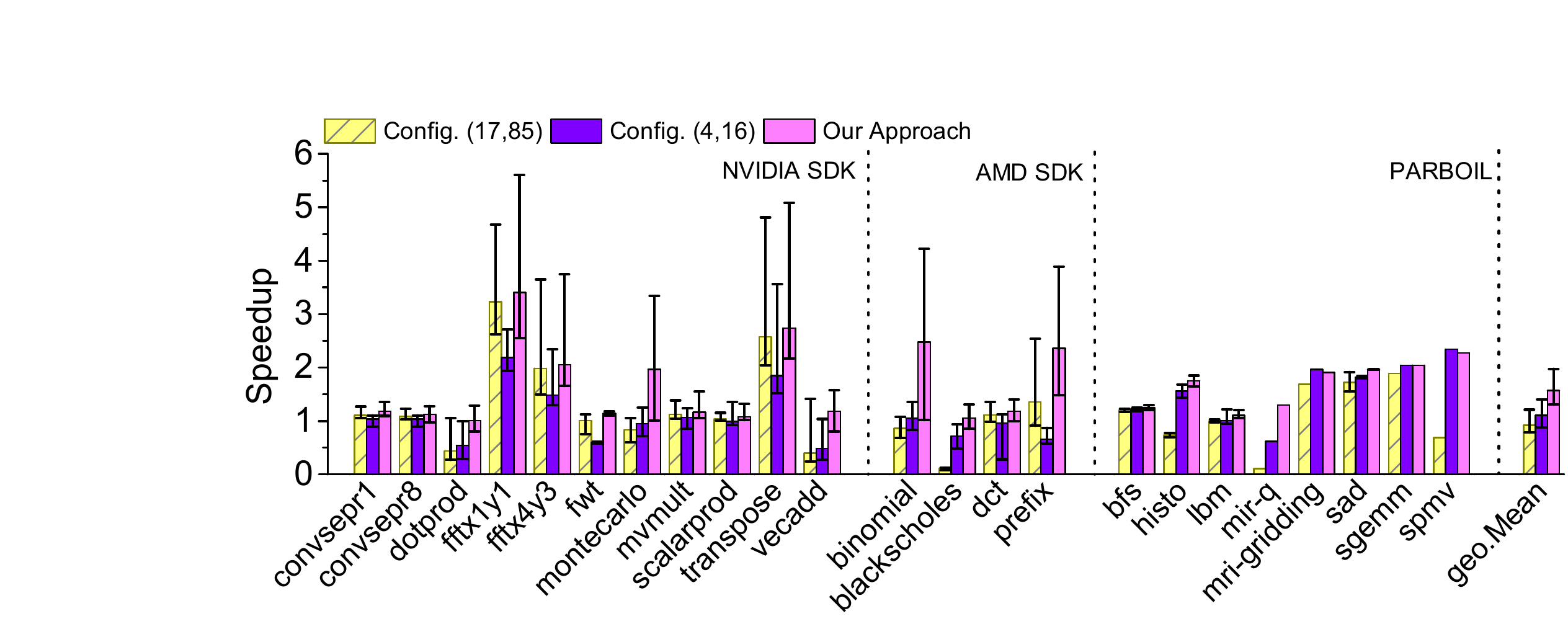}\\
  \vspace{-2mm}
  \caption{Comparing the performance with two fixed configurations: config. $(4,16)$ of 4 partitions and 16 tasks per partition, and config. $(17,85)$ of 17 partitions and 5 tasks per partition.}\label{fig_fixed_config}
\end{figure}

\subsection{Compare to Fixed Stream Configurations}
A natural question to ask is that: \textit{is there a fixed stream configuration that gives reasonable good performance across benchmarks
and datasets?} To answer this question, we compare our predictive modeling based approach to two specific configurations. Our justification
for using the two configurations are described as follows. Our initial results in Section~\ref{sec_mlstream_motivation} indicate that using
the stream configuration of $(4,16)$, i.e. partitioning the cores to 4 groups and running 4 tasks on each partition (16 tasks in total),
gives good performance. The statistics obtained from the training data suggest that the configuration of $(17,85)$ give the best averaged
performance across training samples.

\begin{figure}[!t]
  \centering

  \includegraphics[width=0.5\textwidth]{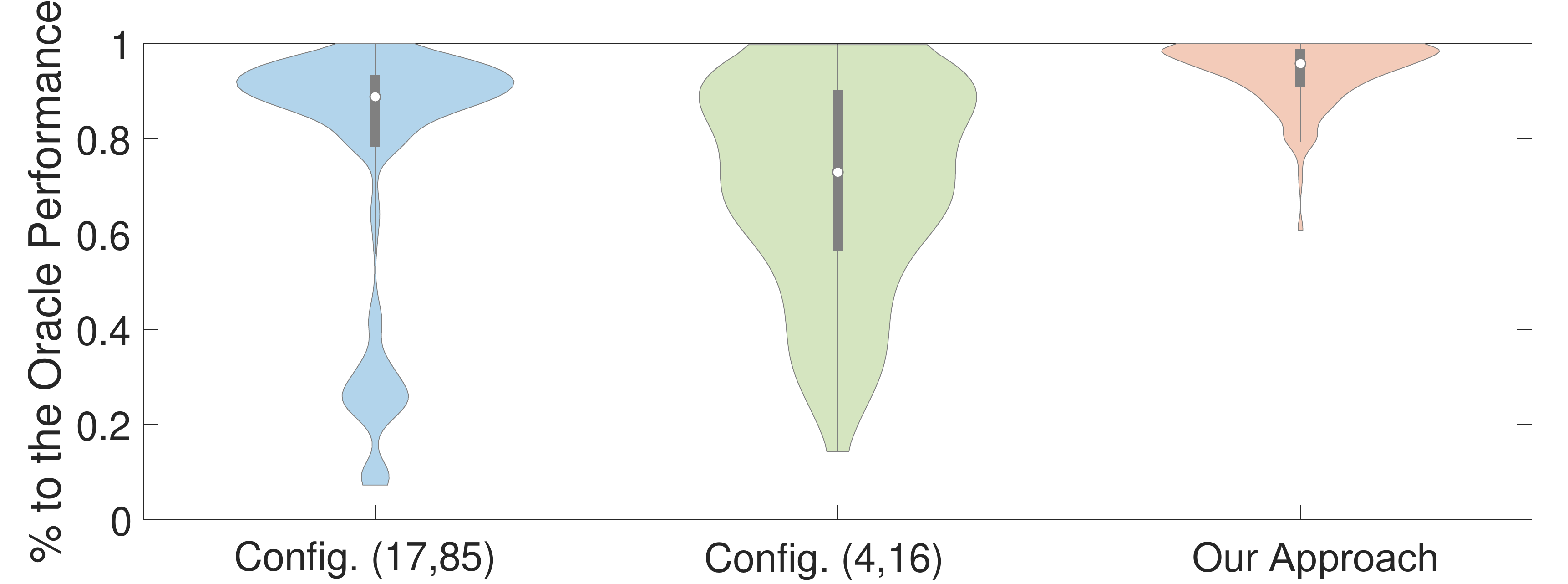}\\
  \caption{Violin plot showing the speedups per scheme across benchmarks and datasets.
The shape of the violin corresponds to the speedup distribution to the oracle performance. The thick black line shows where 50\% of the
data lies.}\label{fig_upper_bound} \vspace{-4mm}
\end{figure}

Based on these two observations, we compare our adaptive approach to two configurations described above. The results are shown in
Figure~\ref{fig_fixed_config}. We observe improved performance for several benchmarks such as \texttt{mri-gridding}, \texttt{transpose},
\texttt{sad}, under both configurations, but slowed down performance for \texttt{dotprod}, \texttt{vecadd}, \texttt{blackscholes},
\texttt{lbm}, and \texttt{mir-q}. For \texttt{prefix}, configuration $(17,85)$ delivers improved performance while configuration $(4,16)$
leads to slowed down performance. Overall, none of the two fixed configurations give an improved performance on average. On average, our
approach outperforms the two fixed configurations by a factor of 1.4, and delivers consistently improved performance across benchmarks and
datasets.

The violin plot in Figure~\ref{fig_upper_bound} shows how far is each of the three schemes to the Oracle performance across benchmarks and
datasets. Our approach not only delivers the closest performance to the Oracle, but also has the largest number of samples whose
performance is next to the Oracle. By contrast, the performance given by the fixed configurations for many samples are further from the
Oracle performance.

This experiment confirms that a fixed configuration fails to deliver improved performance across applications and datasets, and selecting a
right stream configuration on a per program, per dataset basis is thus required.


\subsection{Compare to Alternated Models\label{sec:alt}}

\begin{figure}[!t]
  \centering
  \includegraphics[width=0.5\textwidth]{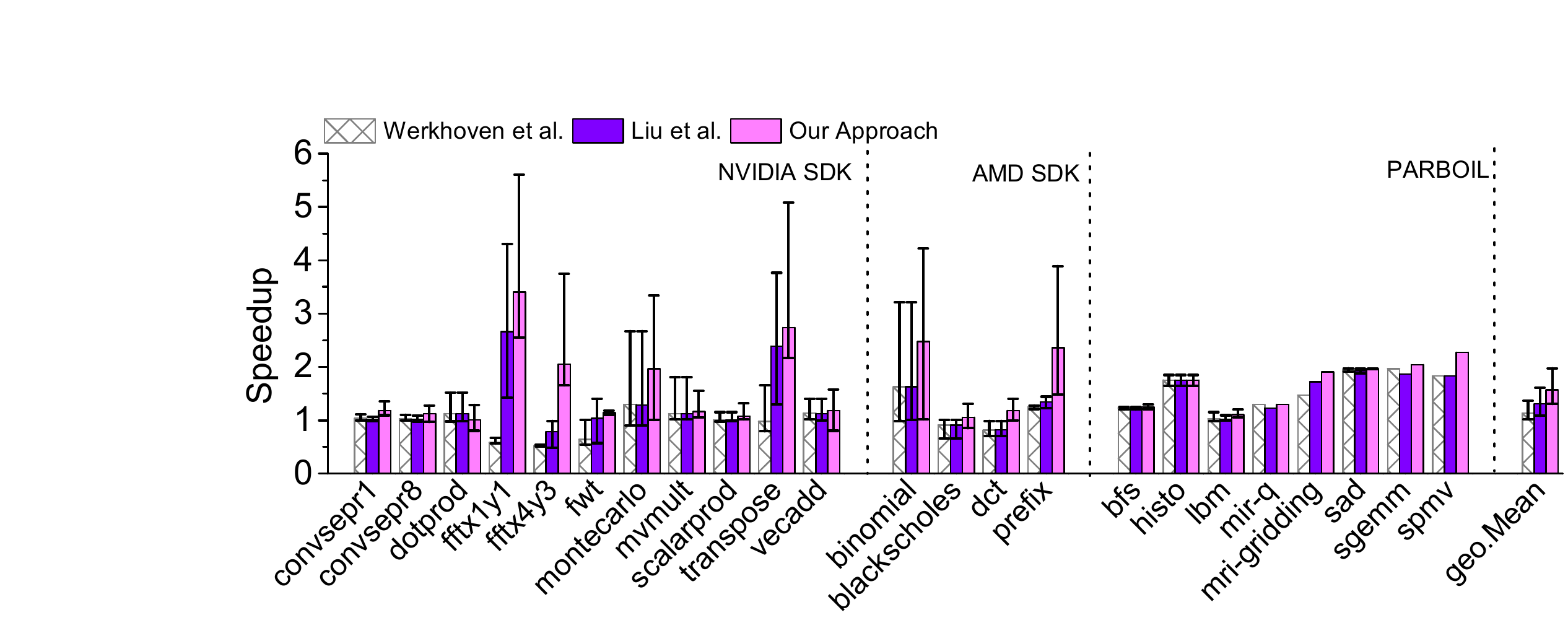}\\
  \caption{Comparing against \Modela and \Modelb
  }\label{fig_hs_model}
\end{figure}

\begin{figure}[!t]
  \centering
  \includegraphics[width=0.5\textwidth]{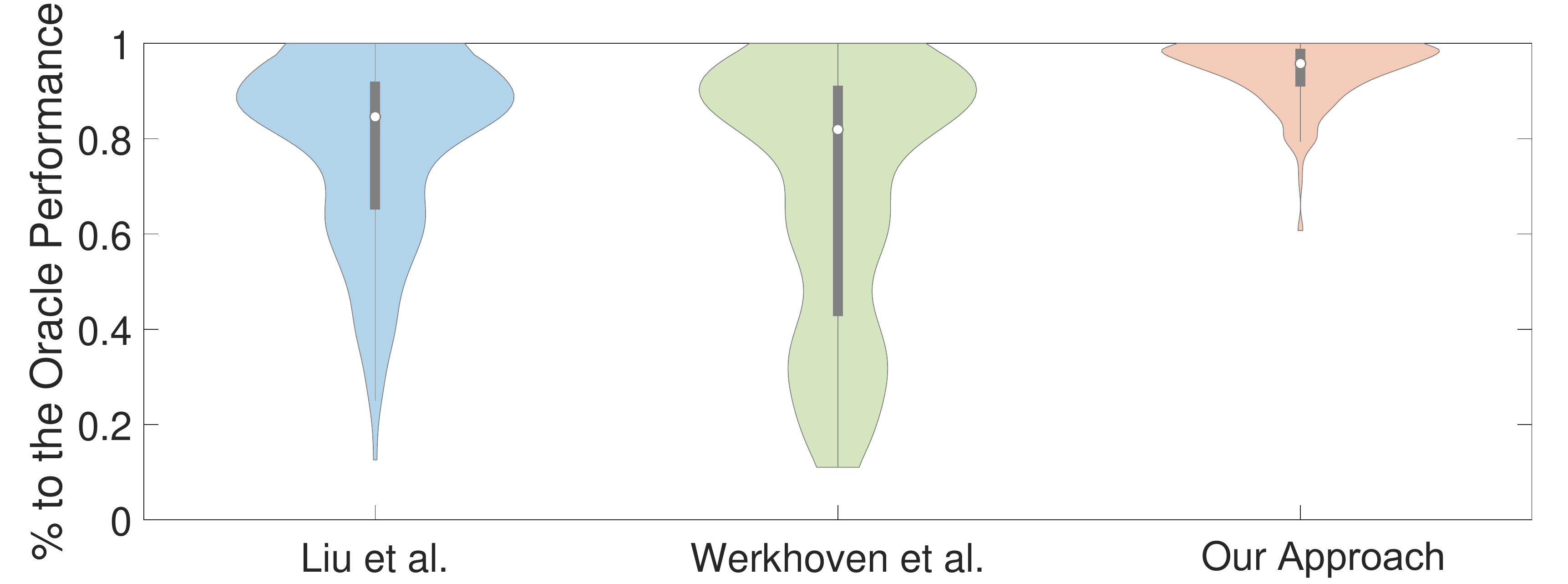}\\
  \caption{Violin plot showing the speedups per scheme across benchmarks and datasets.
The shape of the violin corresponds to the speedup distribution to the oracle performance. The thick
black line shows where 50\% of the data lies.}\label{fig_vl_model}
\end{figure}

In this experiment, we compare our approach to the two recent analytical models
described in Section~\ref{sec:compa}.
The results are shown in Figures~\ref{fig_hs_model} and~\ref{fig_vl_model}.
Both models prefer using $2$ tasks across benchmarks and datasets. This is because that the analytical models simply assume that
task partition has no effect on kernel's performance, and do not consider
the thread management overhead.

From Figure~\ref{fig_hs_model}, we see that our approach can obtain better performance for nearly all programs. For the remaining handful
programs, all three approaches  deliver comparable performance. Compare to Figure~\ref{fig_fixed_config}, we can find the performance of
the analytical models is similar to fixed stream configurations. This is because the performance of the seven programs, such as
\texttt{binomial}, changes dramatically with different stream configurations (see also Figure~\ref{fig_motivation_overall}). The
performance of the remaining programs is not sensitive to the variation of stream configurations. From Figure~\ref{fig_vl_model}, we can
further see that \Modela and \Modelb deliver a speedup within a range on 20\% to 80\%, while the performance of our approach is
centralized on a range between 80\% to 100\%. Thus, our approach delivers consistently better performance compared with the alternative
models.

\subsection{Model Analysis}

\subsubsection{Evaluate the Label Merging Algorithm\label{sec:dist}}


To evaluate our label merging algorithm, we first use the 101 raw labels to train a predictive model.
With the help of the label merging algorithm, we reduce the number of classes to be 28.
Then we use the new labels to train a new predictor and compare the performance of these two models.

\begin{figure}[!t]
  \centering
  \includegraphics[width=0.5\textwidth]{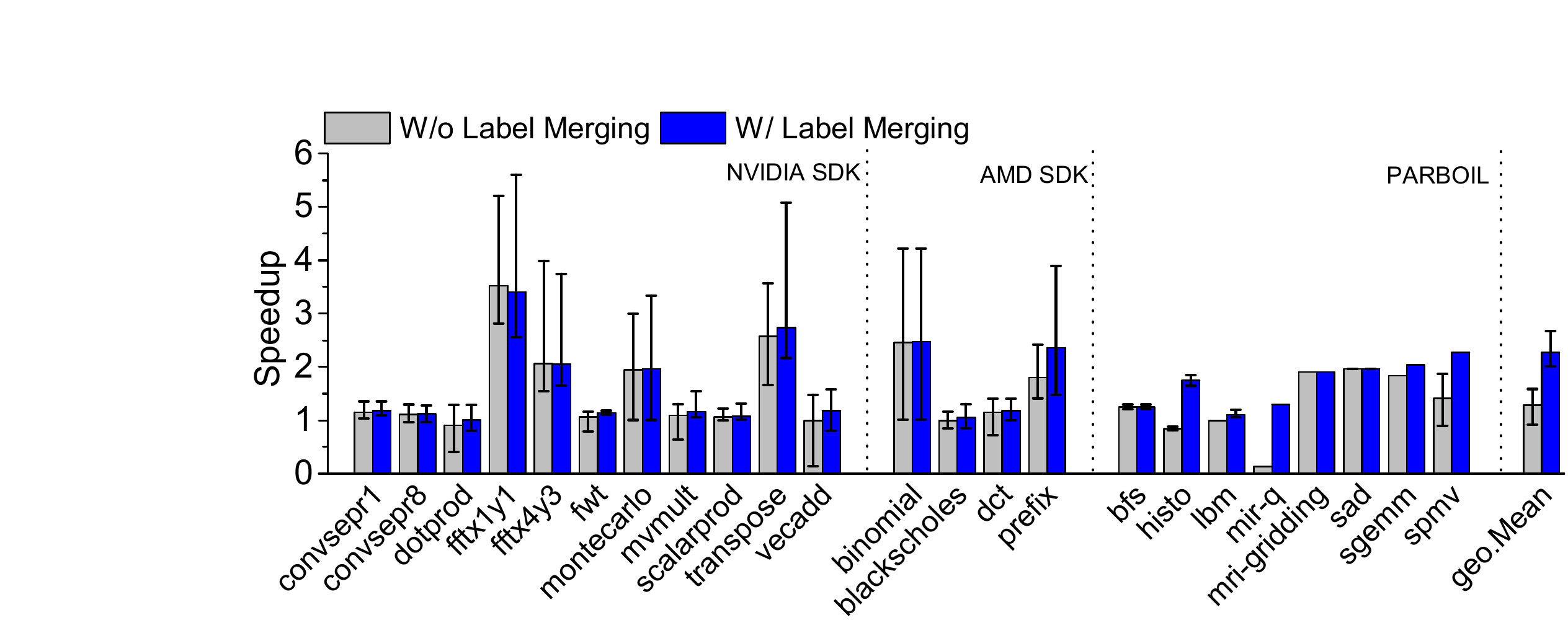}\\
  \caption{Resultant performance with and without label merging.}\label{fig_label_merge}
\end{figure}

We show the result in Figure~\ref{fig_label_merge}. We find that with the labels merge algorithm, the new predictive model performs, on
average, 21\% better than the one without label merging. It indicates that our label merging algorithm can lead to a better predictive
performance by better balance training samples per stream configuration.

\subsubsection{Compare to Alternative Learning Techniques} \label{sec_compare_learning_techniques}
Table~\ref{tbl_learning_technique} shows the average speedup achieved by different machine learning techniques. For each technique, we
follow the same training methodology and use the same features and training examples to build a model. These schemes are implemented using
scikit-learn~\cite{scikitlearn} except for the \SVM models which are built upon libSVM as our approach. We have performed parameter search
on the training dataset to find the best performing model parameters. Specifically, we vary $k$ between 1 and 10 for \KNN models and try
different number of hidden layers and neurons for the \ANN model.

Thanks to the high quality features, all models achieve similar performance (over 1.3x). Our \SVM model based on a \texttt{quadratic}
kernel function gives the best overall performance. This is because the kernel function we used can model both linear and non-linear
relation between the features and the desired labels; and as a result, it predicts the best stream configuration more accurate than other
alternative models. It is to note that the performance of the \ANN model can be further improved if there are more training examples (e.g.,
through synthetic benchmark generation~\cite{Cummins:2017:SBP:3049832.3049843}) and our approach can be used with an \ANN model without
changing the learning process.


\begin{table}[!t]
\caption{Compare to alternative learning techniques.}

\begin{center}
\scriptsize
\begin{tabular}{lclc}
\toprule
\textbf{Learning techniques} & \textbf{Avg. speedup} & \textbf{Learning techniques} & \textbf{Avg. speedup} \\
\midrule
\rowcolor{Gray}  Gaussian \SVM  & 1.37 & Decision tree	 & 1.39 \\
 Backpropagation \ANN	& 1.42 & Linear discriminant	& 1.40 \\
\rowcolor{Gray} Linear \SVM	& 1.43 & Ensemble \KNN	& 1.44 \\
Weighted \KNN	& 1.45  & Our approach	 & 1.57 \\

\bottomrule
\end{tabular}
\end{center}
\label{tbl_learning_technique}
\end{table}

%
%
%
%

\section{Related Work} \label{sec_mlstream_relatedwork}

Our work lies in the interaction of various areas: work partitioning, stream modeling, and predictive modeling.

\cparagraph{Workload Partition.} There is an extensive body of research work in distributing work across heterogeneous processors to
utilize the computation resources to make program run faster~\cite{citeulike:13920415, Luk:2009:QEP:1669112.1669121}.
Prior work in the area typically assumes that the processor configuration is fixed and rely on the operating system to schedule parallel
tasks across parallel processing units. Recent studies show that by partitioning the processing units into groups it is possible to
significantly improve the application performance by overlapping the host-device communication and computation on coprocessors like Intel
XeonPhi~\cite{DBLP:journals/ppl/FangZLTCCY16,DBLP:conf/ipps/NewburnBWCPDSBL16}. However, existing approaches typically rely on manual
tuning to find the processor partition and the best number of streams to run within a partition. As a result, previous approaches cannot
adapt to the change of program behavior due to the change of program inputs. As a departure from prior work, this work develops an
automatic approach to dynamically adjust the processor partition and task-granularity during runtime, considering the characteristics of
applications and input datasets. As a result, our approach can adapt to the change of program behavior and runtime inputs.

\cparagraph{Multiple Streams Modeling.} Gomez-Luna \etal~\cite{citeulike:9715521} develop a set of models to estimate the asynchronous data
transfer overhead on different GPU architectures. The models can be used to estimate the optimal number of streams to use on a given GPU
platform. Werkhoven \etal~\cite{citeulike:13920334} present an analytical model to determine  when to apply an overlapping method on GPUs.
Liu \etal~\cite{citeulike:13920353} also develop an analytical based approach to determine the optimal number of streams to use on GPUs.
However, none of these approaches considers the processor partition. As we have shown in Section~\ref{sec:alt}, ignoring the processor
partitioning parameter can lead to poor performance on Intel XeonPhi. Furthermore, these hand-crafted models have the drawback of being not
portable across architectures as the model is tightly coupled to a specific GPU architecture. Our work advances prior work by employing
machine learning to automatically learn the optimal processor partition and the number of streams/tasks to use. Since our models are
automatically learned from empirical observations, one can easily re-learn a model for a new architecture.

\cparagraph{Predictive Modeling.} Recent studies have shown that machine learning based predictive modeling is effective in code
optimization~\cite{wang2014exploitation,wang2014integrating}, tuning compiler
heuristics~\cite{ogilvie2014fast,cummins2017end,ogilvie2017minimizing}, parallelism
mapping~\cite{Tournavitis:2009:THA:1542476.1542496,Wang:2009:MPM:1504176.1504189,wang2010partitioning,grewe2013portable,wang2013using,DBLP:journals/taco/WangGO14,taylor2017adaptive},
and task
scheduling~\cite{grewe2011workload,emani2013smart,grewe2013opencl,Delimitrou:2014:QRQ:2541940.2541941,ren2017optimise,marco2017improving}.
Its great advantage is its ability to adapt to changing platforms as it has no a prior assumption about their behavior. The work presented
by Wen \etal~\cite{wen2014smart} employs \SVMs to develop a binary classifier to predict that if a given OpenCL kernel can achieve a high
speed up or not. Our work differs from~\cite{wen2014smart} in that it targets a different architecture and programming model, and it
predicts from a larger number of configurations instead of making a binary prediction. We stress that no work so far has used predictive
modeling to model the optimal processor partition and task-granularity on heterogeneous
processors. 

\section{Conclusion} \label{sec_mlstream_conclusion}
This paper has presented an automatic approach to exploit heterogenous streams on heterogenous many-core architectures. Central to our
approach is a machine learning based approach that predicts the optimal processor core partition and parallel task granularity. The
prediction is based on a set of code and runtime features of the program. Our model is built and trained off-line and is fully automatic.
We evaluate our approach on a CPU-XeonPhi mixed heterogenous platform using a set of representative benchmarks. Experimental results show
that our approach delivers, on average, a 1.6x speedup over a single-stream execution. This translates to 94.5\% of the performance given by
an ideal predictor.

\section*{Acknowledgment}
The authors would like to thank the anonymous reviewers for their constructive feedback. This work was partially funded by the National Key
R\&D Program of China under Grant No. 2017YFB0202003, the National Natural Science Foundation of China under Grant No.61602501; the UK
Engineering and Physical Sciences Research Council under grants EP/M01567X/1 (SANDeRs) and EP/M015793/1 (DIVIDEND); and the Royal Society
International Collaboration Grant (IE161012). For any correspondence, please contact Jianbin Fang (Email: j.fang@nudt.edu.cn)

\bibliographystyle{IEEEtran}
\bibliography{IEEEabrv,mybib}

\end{document}